  \documentstyle[12pt,preprint,aps]{revtex}
\include{epsf}

\begin{document}

 \title{Results from the LSND Neutrino Oscillation Search
    for $\rm\overline{\nu}_\mu\longrightarrow\overline{\nu}_e$}
 \author{James E. Hill}
 \address{University of Pennsylvania}
 \date{\today}
\maketitle
\begin{abstract}
      The Liquid Scintillator Neutrino Detector (LSND) at the
Los Alamos Meson Physics Facility sets bounds on neutrino oscillations
in the appearance
 channel \mbox{$\overline{\nu}_\mu \rightarrow \overline{\nu}_{\rm e}$}
by searching for the signature of the reaction
$\rm\overline{\nu}_{e}p\rightarrow{e}^+n:$
an e$^+$ followed by  a 2.2MeV gamma ray from neutron capture.
Five \mbox{$\rm{e}^\pm$--$\gamma$} coincidences
are observed in time with the LAMPF beam, with an estimated
background of 6.2 events.  The 90\% confidence limits obtained are:
 $\Delta \rm m^2 < 0.07eV^2$ for $\sin ^2 2\theta = 1,$ and
$\sin ^2 2\theta < 6\cdot 10^{-3}$ for $\rm\Delta m^2\gtrsim 20\rm eV^2.$
\end{abstract}

{

      The phenomenon of neutrino oscillations \cite{Ponte}
 is a sensitive probe of finite masses in the neutral lepton sector,
the only remaining elementary fermions with unknown masses.
With a few simplifying assumptions, the probability, P, of oscillations
$\rm\overline{\nu}_\mu\longrightarrow\overline{\nu}_e$
can be written
\mbox{$\rm{P}_{\overline{\nu}_\mu\rightarrow\overline{\nu}_e}
            =\sin^2(2\theta)\sin^2\left\{1.27\Delta (m^2(eV^2))
                           L(m)/E(MeV)\right\}$.}
Experiments at accelerators may probe very low, (${\cal O}(10^{-3})$)
values of the mixing strength, $\sin^22\theta$, and mass splittings
$\rm{\Delta m^2}
     \approx{| m_{\overline{\nu}_\mu}^2 -m_{\overline{\nu}_e}^2 | }
     \gtrsim{\cal{O}}(10^{-2}eV^2)$ by searches for
appearance of $\rm\overline{\nu}_e$
in a high intensity $\overline{\nu}_\mu$ beam.

The Liquid Scintillator Neutrino Detector (LSND) was built
at the Los Alamos Meson Physics Facility (LAMPF) in part to
search for oscillations
 $\rm\overline{\nu}_\mu\longrightarrow\overline{\nu}_e.$
The $\rm\overline{\nu}_e$ is detected via the weak charged current interaction
$\rm\overline{\nu}_{e}p\rightarrow{e}^+n.$ For $\rm\overline{\nu}_e$
resulting from the $\overline{\nu}_\mu$
from muon decay at rest, this interaction will produce
a continuous positron spectrum up to 50MeV, which
can be loosely tagged by a delayed time coincidence with
the $\rm 2.2MeV\ \gamma$ ray emitted
after neutron capture on hydrogen in the detector mineral oil.
Since the detector is insensitive to the charge of a particle,
the search is performed above the 35MeV endpoint of the abundant reaction
$\rm{\nu}_e\ ^{12}C\rightarrow e^-\ ^{12}N$.

      LAMPF produces neutrinos from the decay of pions and muons.
The predominant processes are $\pi^+\rightarrow\mu^+\nu_\mu$ and
$\mu^+\rightarrow\rm{e}^+\overline{\nu}_\mu\nu_e$, the decay of
negative particles being suppressed by the high probability
of nuclear capture. Measured pion cross sections, \cite{E866} are used in
the neutrino beam simulation, \cite{RBMC} which predicts
a total flux of $\overline{\nu}_\mu$ from $\mu$ decay at rest
of $3\cdot 10^{13}\rm\overline{\nu}_\mu /cm^2$ at the center of LSND
for the 3.5 months of data analyzed,
with an estimated absolute uncertainty of 7\% and
$\rm{\Phi_{\overline{\nu}_e}}/{\Phi_{\overline{\nu}_\mu}}
 	 	 	 	 	\approx 4\cdot 10^{-4}.$
The neutrino beam also allows a search for $\rm\nu_\mu\longrightarrow\nu_e$
oscillations \cite{Stross} and $\nu_\mu$ quasielastic interactions
\cite{Mike,Us} because a few percent of $\pi^+$ decay in flight
giving a directed beam of higher momentum $\nu_\mu$.
Data for that search are not analyzed here, but the high
energy $\nu_\mu$ contribute to a potential background source.

      The detector is an approximately cylindrical tank
8.3m long and 5.7m in diameter, 29.8m from the beam
production point. Inside the tank, 1220 8'' Hamamatsu phototubes
(PMT) provide 25\% uniform areal
coverage of $\rm 157m^3$ of mineral oil doped with a small
amount, 0.031~g/l, of butyl-PBD scintillator.
The tank is surrounded except on the bottom by
a high light output liquid scintillator veto shield \cite{E645}
viewed by 292 5'' PMT which detects high energy cosmic ray muons
with a measured inefficiency of $\sim 2\cdot 10^{-5}.$
The entire apparatus is under $\rm\sim 2000g/cm^2$ of passive shielding.
A sample of ${\cal O}(10^6)$ electrons from cosmic ray
muon decay is used to determine
the energy response and resolution of the detector, as
well as the efficiency of e$^\pm$ identification.
The fractional energy resolution for electrons at 53MeV is 8\%{,}
and at 35MeV is 10\%{.}
A lower cut on energy of 37MeV is applied to $\rm{e}^\pm$ signal candidates.

      Triggering decisions are based on global sums over 200ns
of detector and of veto shield PMT signal multiplicities.
Events with more than 5 veto PMT signals are vetoed
and initiate a veto of future events for 15.2$\mu$s.
If not vetoed, any event passing a threshold corresponding
to 5MeV electron energy in the detector is designated a primary trigger,
and recorded with up to four events in the previous $52\rm\mu{s}$
that passed a threshold of 17 detector, or 5 veto PMT signals.
Unvetoed primaries with $\ge 125$ detector PMT signals initiate
a lowering of the primary threshold to 21 detector PMT
for the next millisecond so that subsequent low energy gamma
radiation can be detected. Event losses due to the veto shield
and to the loss of events overwritten in memory
total $19\pm 3\%$ for primaries,  and 22\% for gamma rays.

      The beam status is not used in trigger decisions, but
recorded with each event. This allows beam-unrelated
backgrounds to be characterized in detail. The number
expected is predictable with good precision because of the 7.6\% ratio
of beam-on time to beam-off time, and the spatial distribution
of this background aids in distinguishing it from the potential signal.
%
      Particle identification is accomplished by detection of
forward directed prompt light in the \v{C}erenkov cone of
relativistic particles, distinguishing them from
heavier charged particles which produce isotropic, more slowly
emitted scintillation light.
Light and heavy particles are distinguished well by
the $\rm\chi^2/d.f.$ for the reconstructed track direction fit,
the $\rm\chi^2/d.f.$ for the fit to the vertex position, and
the fraction of PMT signals occurring (corrected for photon time of flight)
more than 12ns later than the reconstructed event time.
This product is corrected for an observed energy dependence
based on studies of the cosmic ray muon decay electron sample.
Selection based on this parameter is
89\% efficient in accepting e$^\pm$ and
99\% efficient in rejecting heavy backgrounds.

      LSND took data for 6 weeks in a preliminary
run in 1993, with a $\overline{\nu}_\mu$
 flux of $9\cdot 10^{12}\rm\nu/cm^2$,
 after which the configuration of the active
veto system was slightly changed, the last of the passive shielding
was added, and the electronics were modified to linearize the energy
response of the detector. This change in the energy response changes the
energy dependence of the particle identification scheme, contributing to the
incompatibility of the data from the early run and the $3.5\times$
larger data set from a 1994 run.
Only 1994 data are used in this analysis.


      The properties of neutron capture gamma rays can be studied
using the large flux of cosmic ray neutrons and the 186$\mu$s capture time
of neutrons in mineral oil.
Requirements are placed on the energy of gamma rays, their time of
occurrence and distance relative to the $\rm{e}^\pm$.
It is required that the reconstructed relative distance
be less than 2.4m (The reconstruction error on
gamma rays is approximately 1m.), the relative time
less than 750$\mu$s ($\approx 4$ capture times), and that the gamma-ray
exhibit at least 26 detector PMT hits.
The efficiency for the detection of a capture gamma-ray from
a neutron coming from a signal event is 60\%{.}
(See Table~\ref{BGTab} for the individual efficiencies.)
The measured probability of an accidental $\rm{e}^\pm$--$\gamma$
coincidence passing the selection criteria is 12\%{.}

      Scatter plots of the projections of the reconstructed positions
of background ({\em i.e. } beam-off) events are shown in Fig.~\ref{Scats}.
There are 1381 $\rm{e}^\pm$ shown in the top plots, and
1961 associated gamma-rays shown directly below them.
No $\rm{e}^\pm$--$\gamma$ coincidence is
required in the data of Fig.~\ref{Scats} (except the 1ms trigger
requirement), and the spatial requirement is only that the reconstructed
$\rm{e}^\pm$ be at least 30cm from the surface defined by
the PMT faces.
The coordinate system
 is defined by taking $\hat{z}$ along the
detector cylindrical axis and $\hat{y}$ vertical;
the beam is along $-0.1\hat{y} + 0.99\hat{z}$.
Both \mbox{Y-X} and \mbox{Y-Z} projections are shown.
The surface at the PMT faces is represented
by the plot frame in the Y-Z projections, and by the thin curve
and part of the frame in the Y-X plots.

      Figure~\ref{ScatsOn} shows the same distributions for all (142)
$\rm{e}^\pm$  in time with the beam that pass all the final
$\rm{e}^\pm$ selection criteria except a later, tighter
fiducial volume requirement, and all (189) gamma rays associated
with these e$^\pm$.
These events are the exact beam-on complement
of those shown in Fig.~\ref{Scats}. Any
$\rm\overline{\nu}_\mu\longrightarrow\overline{\nu}_e$ oscillation
signal must emerge from the events in Fig.~2, after the background
indicated in Fig.~1 is properly subtracted, and an e$^\pm$--$\gamma$
coincidence required.

      The inhomogeneity of the background in Fig.~1 and of the potential
signal in Fig.~2 requires confining the fiducial
volume to a region of the detector which is not only more background free,
but within which
there are no strong gradients of event density.
Since the backgrounds for both $\rm{e}^\pm$ and for
coincident $\gamma$ are inhomogeneous, and both enhanced at the bottom
of the detector, the distribution
of distance between primaries and accidentally coincident gamma rays
will not be constant throughout the detector. This problem is
addressed both by tightening the region analyzed, and by the
requirement that coincidences pass each of the separate criteria
on e$^\pm$--$\gamma$ relative time and distance, and gamma ray energy.
A region is
chosen, in part from the data in Figs.~1 and~2, and in part from
separate measurements
of the spatial dependence of e$^\pm$ detection efficiency,
which limits the search to approximately $59\rm m^3$ of active volume.
The volume shown in Figs.~1 and~2 is reduced by 3/4 by
moving inward from 30cm to 50cm from the PMT faces, and
 by a similar factor by excluding 1m at the bottom of the
remaining region.

      Figure~3a shows the $\rm{e}^\pm$ positions for the subset
of the beam-on events in Fig.~2 with a nominally coincident 2.2MeV
gamma-ray.  All of the events shown in Fig.~2 that satisfy the
$\rm{e}^\pm$--$\gamma$ coincidence requirement are included in Fig.~3a.
Within the fiducial volume 
(dotted line) five events remain as possible signal candidates.
Four events are concentrated just outside the periphery of the
fiducial volume, and 16 near 
the bottom of the detector
in 1/3 of the total volume analyzed.

      Figure~3b shows the energy spectrum of the events in Fig.~3a;
the five events within the fiducial volume are in the shaded portion
of the plot.
%
     43 events pass all the same requirements while the
beam is off, implying a background from beam-unrelated sources of
$\rm (43\times 0.076 = )3.3\pm  0.49(stat)\pm 0.04(sys.)$ events.
%
Almost as large in number is the class of events with neutrino
induced e$^-$ in the correct energy range
in accidental coincidence with low energy gamma radiation
in the detector.
These include \mbox{$\rm\nu_\mu e\rightarrow\nu_\mu e$} and
 \mbox{$\rm\nu_e e\rightarrow\nu_e e$,}
$\rm{\nu}_e\ ^{13}C\rightarrow e^-\ ^{13}N$,
with an endpoint of about 50MeV, and
$\rm{\nu}_\mu\ ^{12}C\rightarrow \mu^-\ X$
(from higher energy neutrinos from pion
decay in flight) where the muon is undetected.
There are also expected to be e$^-$ from
\mbox{$\rm{\nu}_e\ ^{12}C\rightarrow e^-\ ^{12}N$} 
 which have energy measured within $1\sigma$ of the reaction
endpoint of 35MeV.
The excess of $\rm e^\pm$ (beam-on - beam-off) satisfying
 all the $\rm{e}^\pm$ selection criteria,
but without a coincident gamma-ray is made up of these events
(without an accidental gamma-ray) plus any beam related
e$^\pm$ missing a real gamma-ray coincidence.
Both the accidental coincidence probability and the beam excess of
neutrino-induced e$^\pm$ are measured. The background
from beam induced e$^\pm$ in accidental coincidence with a gamma ray
is calculated from these measurements as $2.5\pm 0.9$.
There is an expected background of events from
$\rm\overline{\nu}_e$ contamination from $\mu^-$ decay in the beam stop
of about 0.3 events.
      Other backgrounds have been calculated, and are all found
to be relatively low.

      These results are stable against variation of the
above selection requirements. For example,
extending the region analyzed to $\rm y=-100cm$, {\em i.e. }
50cm lower ($\approx 1.2$ times
the fiducial volume above) leaves 7 beam-on events with a
background of 8.7 events. Similarly,
tightening the requirement on the e--$\gamma$ relative distance to 1m
(from 2.4m) leaves only 2 beam-on events with a background of 1.5,
or, setting the limit on relative time to 375 (from $750\mu$s)
leaves 1 event with a background of $3.5$.

      Major backgrounds are summarized
in Table~\ref{BGTab}. The total background of 6.2 events
leaves no apparent signal for
$\rm\overline{\nu}_\mu\longrightarrow\overline{\nu}_e$ oscillations.
The resulting limits on the oscillation parameters
$\Delta {\rm m}^2$ and $\sin ^2 2\theta$
based on this number of events are shown in Fig.~\ref{Limits}.
The 90\% confidence limit is $\sin ^2 2\theta < 6\cdot 10^{-3}$ for
$\Delta\rm m^2 \gtrsim 20eV^2$ and
$\Delta\rm m^2 <\rm 7\cdot 10^{-2}eV^2$ at $\sin ^2 2\theta = 1.$
Ignoring all beam-related backgrounds gives the
high $\rm\Delta m^2$ 90\% C.L.: $\sin ^2 2\theta < 8\cdot 10^{-3}.$

      Study of the data for the search for $\rm\nu_\mu\longrightarrow\nu_e$
oscillations from the LAMPF pion decay in flight neutrino beam
\cite{Stross,Mike} is in progress.

\acknowledgements
 The author would like to acknowledge the support and many fruitful
discussions with A.K.Mann. I would also like to thank the members of the
LSND collaboration and the students who worked on the project,
with whom most of this work was done,
and the staff of LAMPF. This work is supported in part by the
U.S. Department of Energy. The material presented here is
part of a Ph.D. thesis submitted to the University of Pennsylvania,
which will be made available for anonymous ftp upon completion of
thesis defense in early May,'95.

%

}

\onecolumn
 \pagebreak

\begin{figure}
\caption{Positions of the beam-unrelated backgrounds. The scatter plots
 	 show positions of individual $\rm e^\pm$ (upper plots) and
 	 associated gamma rays (lower plots)
 	 in YX and YZ projection. No e--$\gamma$ coincidence is required for
 	 these samples, beside the 1ms time coincidence required by
 	the on-line trigger. The tangent surface to the PMT faces is
 	the plot boundary in the Y--Z projection, and the solid arcs together
 	with the plot boundary in the Y--X projection. $\rm e^\pm$ are
 	restricted to the region $\rm d_e^{PMT}>30cm$.
 	The distance between centers of adjacent PMT is 35cm.
        \label{Scats} } 
\end{figure}
 \vspace*{-5mm}
\begin{figure}
\caption{Positions of the beam-on events. The 142 $\rm e^\pm$ shown
 	(upper plots) are the superset from which all e--$\gamma$
 	coincidences must be drawn. The lower plots show the positions
 	of the 189 gamma rays that occur in the 1ms windows these electrons
 	have initiated. The The neutrino beam is along
 	$-0.1\hat{y} + 0.99\hat{z}.$
 	These data represent the full 4 months (5904C) of beam.
 	\label{ScatsOn} }
\end{figure}


\begin{figure}\caption{(a)The 25 beam-on e--$\gamma$
 	coincidences, before the application of the fiducial volume cut.
 	The fiducial region, indicated by the dotted line, is 50cm
 	from the PMT faces, except at the bottom of the detector, where
 	it is 149cm from the PMT faces. Events within this volume are
 	denoted as solid circles, while those outside are represented as
 	open circles. This volume is 55\% of the volume represented
 	in Figs.~1 and~2.
 	(b) The e$^\pm$ spectrum for these 25 events. The five events
 	passing the fiducial volume cut make up the shaded protion of
 	the plot.\label{OnSet} }
\begin{tabular}{c}
\end{tabular}
\end{figure}
\begin{figure}\caption{Exclusion plot for mixing parameters
 	       $\Delta {\rm m}^2$ and $\sin^2 2\theta$.
               The region to the right of and above the solid curves is
	       excluded by this analysis at the
               90(95)\% confidence level.
               At high $\rm\Delta (m^2)$, $\sin^22\Theta < 6\cdot 10^{-3} $
               and at full mixing $\rm\Delta (m^2) < 0.07eV^2$
 	       (90\% C.L.). 
 	       The dash-dot curve is the result of BNL-E776,
 	       and the dashed curve is the result from KARMEN.
 	       \label{Limits}}
\end{figure}

{ \begin{twocolumn}
\begin{table}\caption{Efficiency of selection for the
final sample.The
 efficiency of in-time veto shield and previous detector activity
 are measured by generating external control events with
 a laser. The cut on previous activity of $\rm{e}^\pm$
 is motivated by the large background of muon decay.\label{darCuts}}
\begin{tabular}{|c|d|}
Cuts applied on & Acceptance \\ \hline
previous activity & 0.65 \\
in-time veto for e$^+$& 0.87 \\
e$^+$ particle ID & 0.89 \\
Live time for $\rm e^\pm$ & 0.81 \\ \hline
$\rm{e^\pm}$--$\gamma$ correlation & 0.93 \\
$\gamma$ energy & 0.9  \\
in-time veto for $\gamma$ & 0.93 \\
Live time for $\gamma$ & 0.78 \\ \hline
Total & 0.25 \\
\end{tabular}
\end{table}

\begin{table}\caption{Expectation values for \label{BGTab}
some important backgrounds. The two largest are
measured (not calculated).}
\begin{tabular}{|c|c|d|}
$\nu$ source &
                reaction in LSND &
                         $\rm N_{expected} $. \\ \hline
 \multicolumn{2}{|c|}{Beam-unrelated background} &
                                $3$.$3\pm 0$.$5$ \\
 \multicolumn{2}{|c|}{Beam induced $\rm{e}^\pm$ with accidental $\gamma$} &
        $2$.$5\pm $0.$9$ \\ \hline
%
%
 $\mu^- \rightarrow \nu_\mu \overline{\nu}_{\rm e} {\rm e} ^- $ &
        $\rm\overline{\nu}_ep\rightarrow e^+n $ &
                0.$3\pm 0$.$1$ \\
 $\pi^-\rightarrow {\rm e^-\overline{\nu}_e} $ &
        $\rm\overline{\nu}_ep\rightarrow e^+n $ &
                 $< 0$.$01$ \\ \hline
%
%
 $\pi \rightarrow \mu{\nu}_\mu $ in flight &
 	$\rm\stackrel{(-)}{\nu}_{\mu}X\rightarrow\mu{nX}$ & 0.1$\pm$0.1\\ \hline
%
\multicolumn{2}{|r||}{\Large Total} & $6$.$2\pm 1$.$6$ \\
\end{tabular}
\end{table}
\end{twocolumn}
}

\begin{references}
\bibitem{Ponte}
 	 Pontecorve, B., {\em Zh. Eksp. Theor. Phys. } {\bf 33}, 549
        1957, translated {\em Sov. Phys. JETP} {\bf 6}, 429 (1958).
 \bibitem{E866}R. C. Allen, {\em et al.,} Nucl. Instr. Methods,
 	{\bf A291}, 347 (1990).
\bibitem{RBMC}
        R. L. Burman, {\em et al.,} Nucl. Instr. Meth.,
        {\bf A291}, 621 (1990).
\bibitem{Stross}
        Strossman, W., {\em A $\rm\nu_\mu\longrightarrow\nu_e$
        Oscillation Search,}
        Ph.D. Thesis, Univ. of California, Riverside, 1995. (unpublished)
\bibitem{Mike}
        Michael Albert, {\em A Measurement of the Reaction
        $\rm C(\nu_\mu ,\mu^- )X $ Near Threshhold}, Ph.D. Thesis,
        Univ. of Penn., 1994. (unpublished)
\bibitem{Us}
 	M. Albert, {\em et al.,} Phys. Rev. C, {\bf 51}, R1065, (1995).
\bibitem{E645}
        J. Napolitano, {\em et al.,} Nucl. Instr. Meth.,
        {\bf A 274}, 152 (1989).
\end{references}
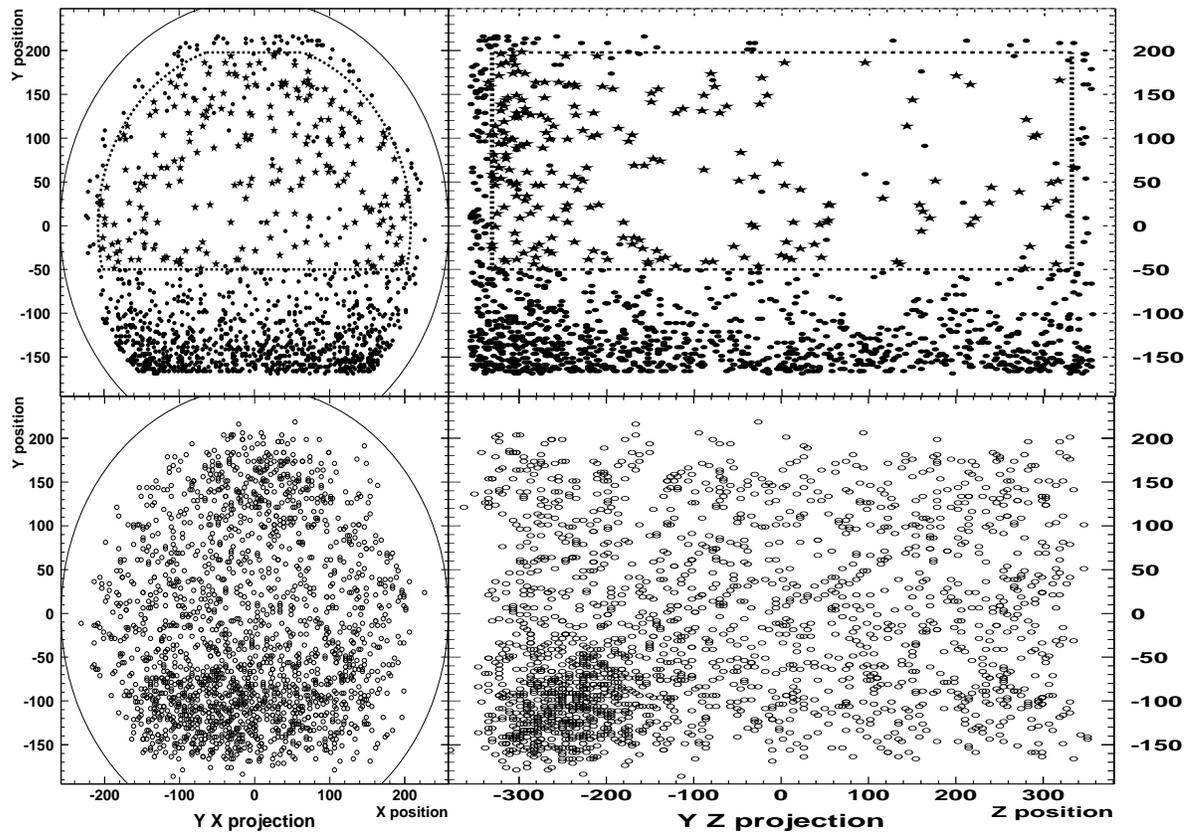
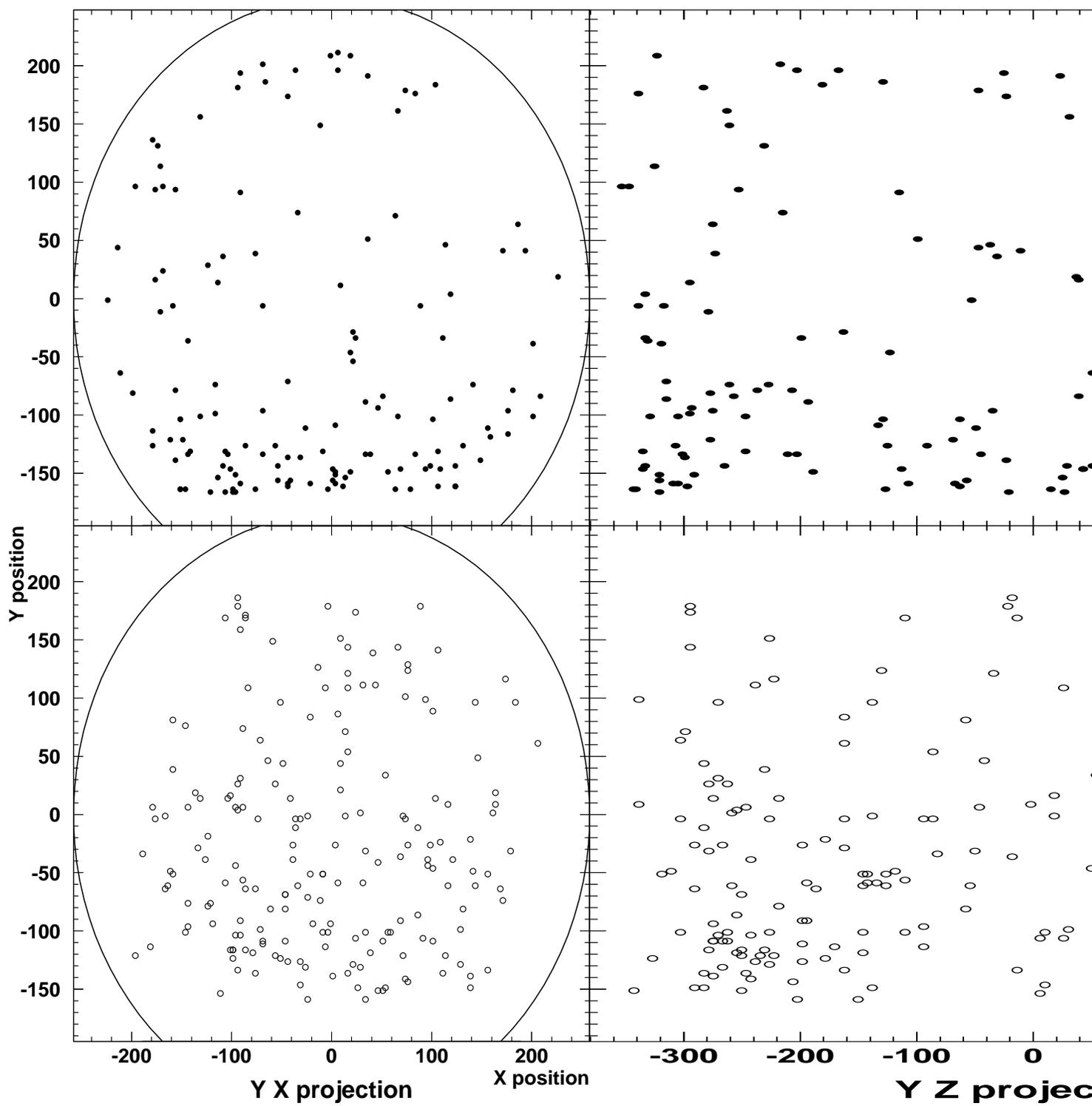
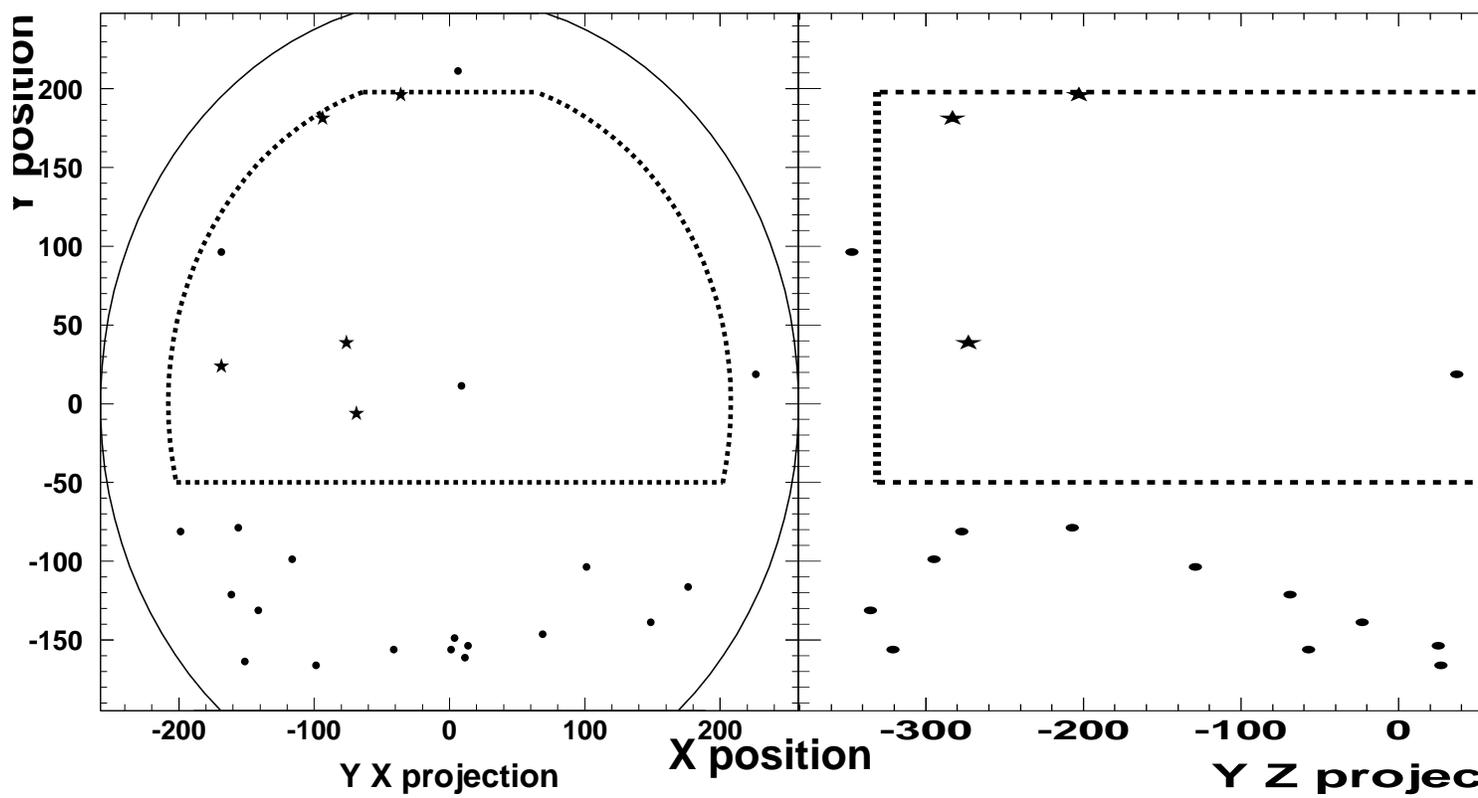
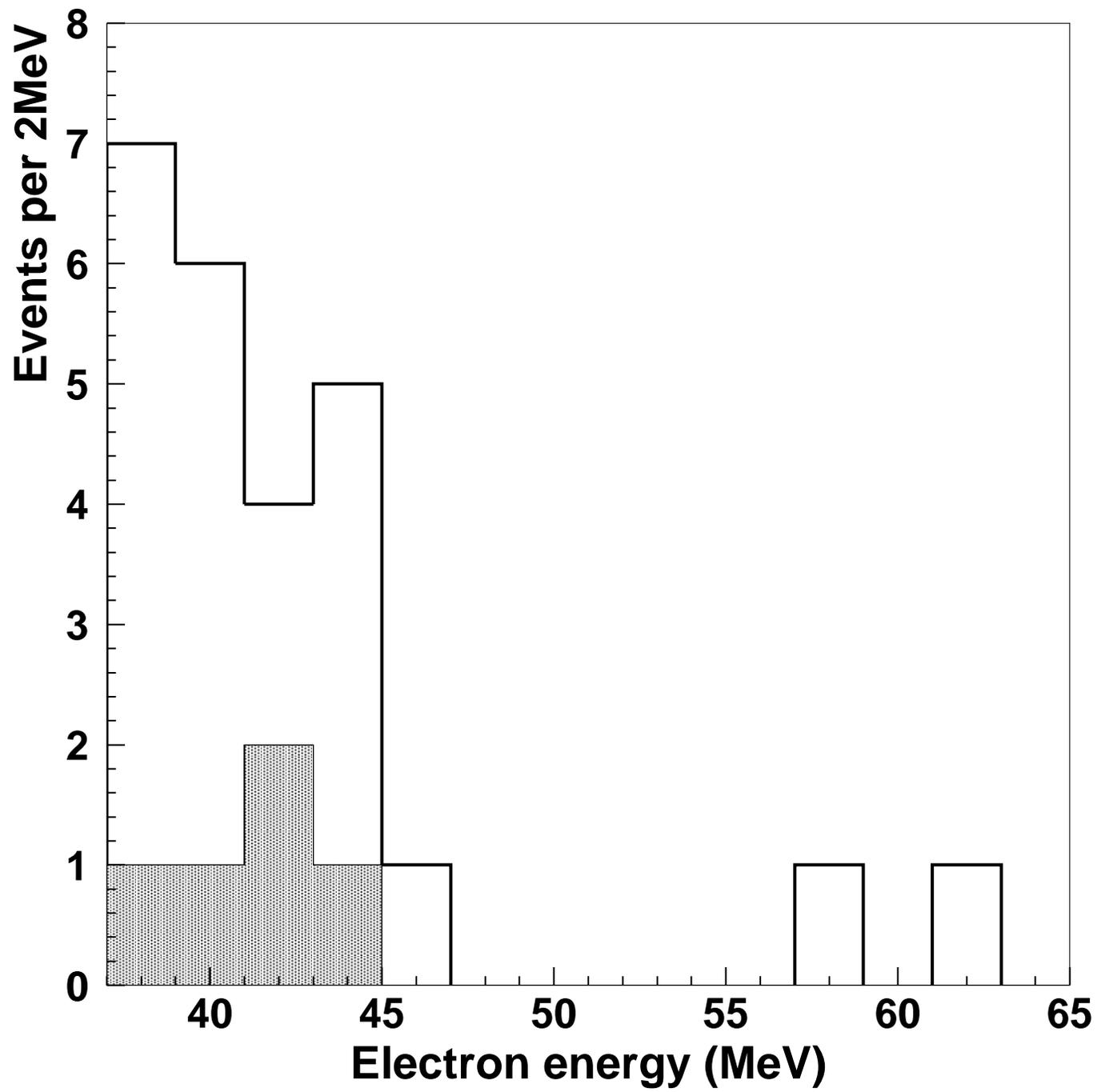
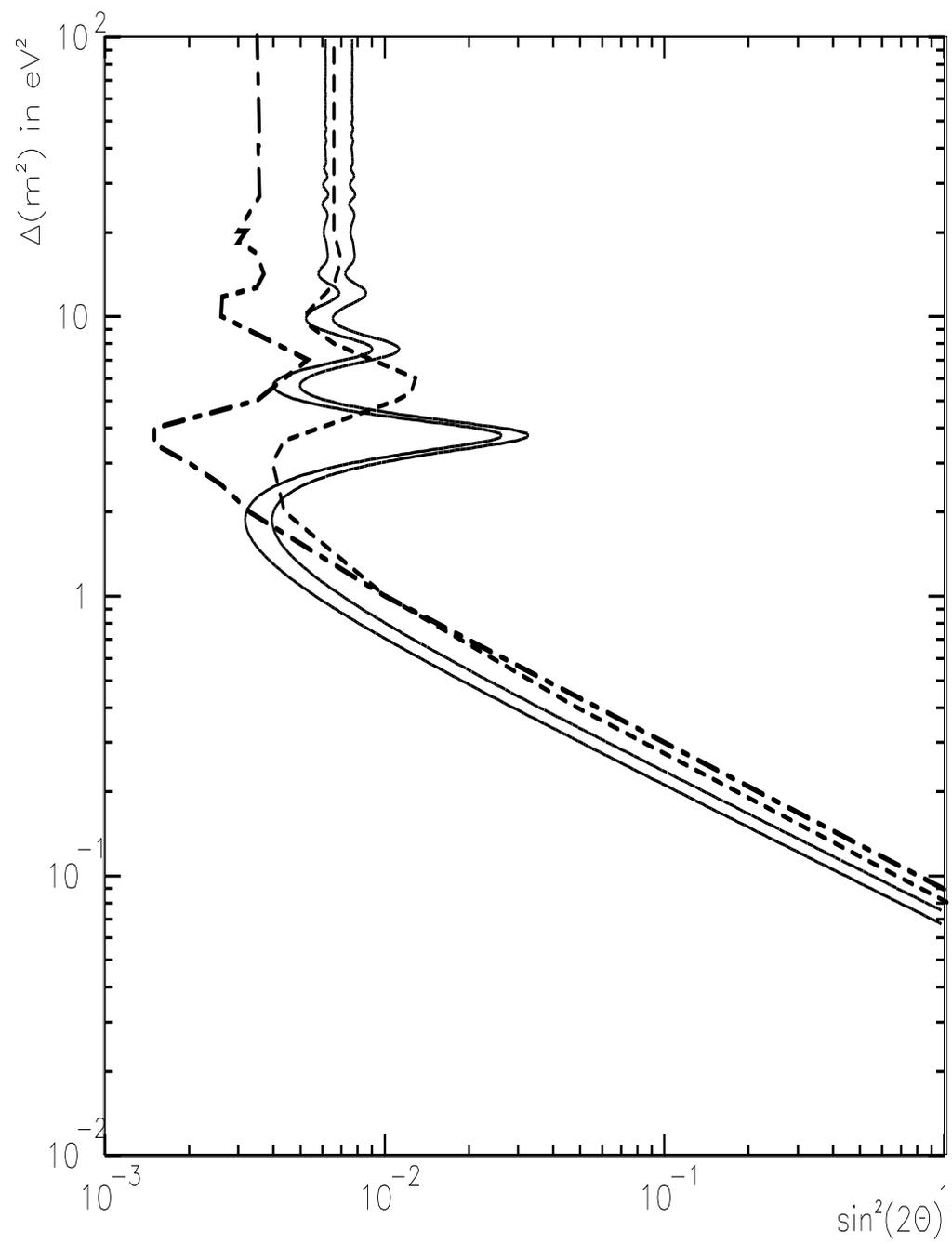
\end{document}